\documentclass[sigconf]{acmart}

\usepackage{graphicx}
\usepackage{url}
\usepackage{amsmath}
\usepackage{booktabs}

\newtheorem{assumption}{Assumption}
\newtheorem{remark}{Remark}

\setcopyright{acmlicensed}
\copyrightyear{2026}
\acmYear{2026}
\acmDOI{XXXXXXX.XXXXXXX}
\acmISBN{978-1-4503-XXXX-X/XXXX/XX}
\acmConference[KDD CJ '26]{5th End-to-End Customer Journey Workshop at KDD 2026}{August 10, 2026}{Jeju Island, South Korea}

\title{Bias-Aware Confidence Intervals for Synthetic Control via Placebo-in-Time Bootstrap}

\author{Song (Vinson) Wei,\ \ Sonia Park,\ \ Niteesh Kalangi,\ \ and\ \ Jason Huang}
\affiliation{%
  \institution{Databricks}
  \country{USA}
}
\renewcommand{\shortauthors}{Wei et al.}

\begin{document}

\begin{abstract}
Synthetic control (SC) methods are among the most widely used tools for causal inference without randomization. The standard Gaussian confidence interval around the estimated effect is simple, fast, and reliably directional when the treatment signal is strong, so practitioners default to it for good reason. Most treated populations, however, are bottom-heavy in intensity, and for them the SC model's systematic bias rivals or exceeds the signal even under good pre-treatment fit. Because this bias shares sign across units it does not average out, and the Gaussian confidence interval shrinks past it and converges on a wrong center. The failure is not imprecision but misdirection: a positive effect estimated as negligible is a missed opportunity, while a negligible effect estimated as significantly positive leads to continued investment in an intervention that is not working. No existing confidence interval for the SC effect measures this bias. We propose a placebo-in-time bootstrap that estimates the bias distribution directly from the observed panel. For each treated unit the procedure backdates the treatment onset and refits the SC model at each placebo onset; the resulting placebo gaps are draws from the same bias distribution that contaminates the real estimate, and bootstrapping them yields a critical value calibrated at the zero null. Because the method resamples realized model error rather than a hypothesized effect, coverage is trajectory-agnostic: it holds at fixed width regardless of how the true effect evolves over time.
\end{abstract}

%\ccsdesc[300]{Mathematics of computing~Nonparametric statistics}
%\ccsdesc[300]{Applied computing~Economics}
%\ccsdesc[100]{Mathematics of computing~Time series analysis}

\maketitle
\begingroup\renewcommand\thefootnote{}\footnotetext{Correspondence to Song (Vinson) Wei (\texttt{vinson.wei@databricks.com}). Song (Vinson) Wei designed the method, built the proof of concept and first production deployment at Databricks, and led the research project. Sonia Park contributed to production deployment scalability. Niteesh Kalangi conducted the numerical simulations during his internship. Jason Huang provided technical and domain-expertise guidance.}\endgroup

%% =========================================================
\section{Introduction}\label{sec:intro}

Synthetic control (SC) methods~\cite{Abadie2003,Abadie2010} estimate the causal effect of an intervention by constructing, for each treated unit, a weighted combination of untreated units that approximates the counterfactual outcome under no treatment. In applications spanning policy evaluation~\cite{Abadie2010}, economics~\cite{Abadie2003}, and industry~\cite{WeiHuang2025,Costa2023}, the confidence interval around the estimated average effect is typically the decision surface: it determines whether an intervention is judged effective, expanded, or discontinued. The standard Gaussian confidence interval is simple, fast, and reliably directional when the treatment signal is strong relative to the model's estimation error, which is why it remains the default in practice.

That default breaks down for the populations that matter most. Most treated populations are bottom-heavy in treatment intensity: incremental interventions, marginal exposures, and late-adopter cohorts concentrate the majority of units at low per-unit effects, where strong responders are rare~\cite{Rogers2003}. For these units the SC model's systematic estimation bias rivals or exceeds the signal, even though pre-treatment fit is good: simplex-constrained weights cannot exactly replicate the treated unit's factor exposure, so a residual gap persists under no treatment~\cite{FermanPinto2021}. Because this bias shares sign across units it does not cancel in the average, and the Gaussian confidence interval, calibrated only for noise, shrinks past it and converges on a wrong center. The result is not imprecision but misdirection: a truly positive effect estimated as negligible represents missed opportunity, while a negligible or negative effect estimated as significantly positive leads to continued investment in an intervention that is not working. In a real-world measurement system over a large cohort~\cite{WeiHuang2025}, either error compounds at scale.

Every existing confidence interval for the SC effect substitutes an assumption for the missing bias rather than estimating it. Variance-based confidence intervals (jackknife, subsampling~\cite{Li2020}) recalibrate dispersion around the contaminated center. The in-space placebo~\cite{Abadie2010} assumes a treated--donor exchangeability that structural mismatch violates~\cite{FermanPinto2021,HahnShi2017}. The closest benchmark, conformal inference~\cite{ChernozhukovWuthrichZhu2021}, requires the analyst to specify how the treatment effect evolves over time and collapses when that specification is wrong, a fragility we demonstrate in Section~\ref{sec:regime}. None measures the bias distribution itself (Appendix~\ref{app:litsurvey}).

We propose a placebo-in-time bootstrap that estimates the bias distribution directly from the observed panel, turning the in-time placebo of Abadie et al.~\cite{Abadie2015} from a binary diagnostic into a quantitative confidence interval for the aggregate SC effect. For each treated unit, the procedure backdates the treatment onset and refits the SC model at each placebo onset; the resulting placebo gaps are draws from the same bias distribution that contaminates the real estimate, and bootstrapping them yields a bias-aware critical value calibrated at the zero null (Section~\ref{sec:method}). Because the resampling target is placebo effects rather than SC weights, the construction sidesteps the known inconsistency of the naive bootstrap for SC estimators~\cite{Li2020}. Its distinguishing property is \emph{trajectory-agnosticism}: coverage holds at fixed width regardless of how the effect evolves over time, exactly where conformal inference, forced to guess that shape, collapses (Section~\ref{sec:regime}). Under correct specification the Gaussian confidence interval remains better and should be kept, and the coverage ceiling is the price of conditioning on a single panel, not a defect of the estimator.

%% =========================================================
\section{Methodology}\label{sec:method}

The idea is one move: refit the SC model at many backdated onsets, and the placebo gaps that come back trace the bias the real estimate suffers, so their bootstrap quantile becomes the critical value. We first set up the gap decomposition and the aggregate target, then make the construction precise.

\subsection{Problem Setup}
Let $Y_{i,t}$ be the observed outcome for treated unit $i \in \{1, \ldots, M\}$ at time $t \in \{1, \ldots, T\}$, and let $T_i^*$ be its treatment onset. The SC estimator gives a counterfactual $\hat{Y}_{i,t}^{(0)}$ for $t \geq T_i^*$, the time-indexed gap is $\hat{\Delta}_{i,t} = Y_{i,t} - \hat{Y}_{i,t}^{(0)}$, and the unit-level gap is $\hat{\Delta}_i = \sum_{t \geq T_i^*} \hat{\Delta}_{i,t}$. This gap decomposes as
\begin{equation}\label{eq:decomposition}
  \hat{\Delta}_i \;=\; \tau_i \;+\; b_i \;+\; \varepsilon_i,
\end{equation}
where $\tau_i$ is the true treatment effect, $b_i$ the systematic SC model bias (the gap that would persist under no treatment when pre-treatment fit is imperfect and assignment is correlated with unobservables~\cite{FermanPinto2021}), and $\varepsilon_i$ observation noise; write $\bar{\Delta}$, $\bar{\tau}$, and $\bar{b}$ for the cross-unit means of the gaps, effects, and biases.

The Gaussian (naive) confidence interval is as follows:
\begin{equation}\label{eq:naive-ci}
  \bar{\Delta} \;\pm\; z_{\alpha/2} \cdot \frac{\hat{\sigma}_{\Delta}}{\sqrt{M}},
\end{equation}
with $\hat{\sigma}_{\Delta}$ the cross-unit standard deviation of the gaps and $z_{\alpha/2}$ the standard normal quantile at level $\alpha$. When the biases share a direction ($\mathbb{E}[b_i] \neq 0$), $\bar{b}$ does not vanish as $M$ grows while $\hat{\sigma}_{\Delta}/\sqrt{M} \to 0$, so the interval shrinks around the contaminated center $\bar{\tau} + \bar{b}$.

When unit sizes vary we normalize and define the aggregate relative effect as:
\begin{equation}\label{eq:aggregate}
  R \;=\; \frac{\sum_{i=1}^{M} \sum_{t=T_i^*}^{T} \hat{\Delta}_{i,t}}
             {\sum_{i=1}^{M} \sum_{t=T_i^*}^{T} Y_{i,t}}.
\end{equation}
The numerator equals $M\bar{\Delta}$, so the bias $\bar{b}$ contaminates $R$ just as it does $\bar{\Delta}$. The coverage target is the realized aggregate relative effect
\begin{equation}\label{eq:rtrue}
  R_{\mathrm{true}} \;=\; \frac{\sum_{i=1}^{M} \tau_i}{\sum_{i=1}^{M} \sum_{t=T_i^*}^{T} Y_{i,t}},
\end{equation}
the value $R$ would take if the SC counterfactual were exact. The construction rests on three conditions, of which the load-bearing one is \emph{bias transportability across onsets} (A2): the bias the SC model makes when fit on a backdated (fictitious) treatment onset is drawn from the same distribution as the bias at the actual onset, so that the placebo gaps are informative about the real error. Formal statements of A1--A3 are in Section~\ref{sec:assumptions}.

\begin{remark}[Calibration scope]\label{rem:target}
$R_{\mathrm{true}}$ is a realized, panel-dependent target, not a fixed population parameter, and ``coverage'' throughout means coverage of $R_{\mathrm{true}}$. The calibration is exact at the zero null ($R_{\mathrm{true}} = 0$), where the placebo pool is uncontaminated. At nonzero effects, treatment leaks into the placebo windows and shifts the pool's location; the symmetric absolute-deviation construction absorbs this shift into width, which is why coverage recovers rather than degrades at high treatment intensity.
\end{remark}

\subsection{Proposed Method}

\paragraph{Placebo-in-time construction.}
For each treated unit $i$ and each backward shift $\delta \in \{1, \ldots, \delta_{\max}\}$, define a fictitious (``placebo'') treatment onset $T_i^{(\delta)} = T_i^* - \delta$ at a time when no treatment was administered, refit the SC model on the data before this placebo onset $\{t < T_i^{(\delta)}\}$, and compute the placebo gap $\hat{\Delta}_{i,t}^{(\delta)} = Y_{i,t} - \hat{Y}_{i,t}^{(\delta,0)}$ for $t \geq T_i^{(\delta)}$, yielding $M \times \delta_{\max}$ placebo trajectories.

\paragraph{Bootstrap over the placebo pool.}
Let $\mathcal{P} = \{(i, \delta)\}$ be the pool. The pool is strongly dependent within unit---a unit's placebo windows share the final $L = T - T_i^* + 1$ post-treatment periods and are fit with nearly identical warm-started weights---so its \emph{effective} size is closer to $M$ than to $M \cdot \delta_{\max}$. Resampling $(i,\delta)$ pairs i.i.d.\ nonetheless tracks the cross-unit conditional dispersion, because near-replicate draws add no spurious dispersion. We partition $\mathcal{P}$ into $K$ strata on pre-treatment covariates so bootstrapped cohorts match the treated composition, and each replicate $m = 1, \ldots, B$ draws $n_k$ pairs with replacement per stratum and computes the placebo aggregate over the sampled multiset $\mathcal{S}^{(m)}$,
\begin{equation}\label{eq:bootstrap}
  R^{(m)} = \frac{\sum_{(i,\delta) \in \mathcal{S}^{(m)}} \sum_{t \geq T_i^{(\delta)}} \hat{\Delta}_{i,t}^{(\delta)}}
             {\sum_{(i,\delta) \in \mathcal{S}^{(m)}} \sum_{t \geq T_i^{(\delta)}} Y_{i,t}}.
\end{equation}

\paragraph{Calibrated-at-zero interval.}
Define the critical value as the $(1-\alpha)$-quantile of the absolute placebo effects and form the symmetric confidence interval,
\begin{equation}\label{eq:ci}
  c_\alpha \;=\; Q_{1-\alpha}\!\left(|R^{(1)}|, \ldots, |R^{(B)}|\right), \qquad
  \bigl[\,R - c_\alpha,\;\; R + c_\alpha\,\bigr],
\end{equation}
rejecting $H_0\!:\,R_{\mathrm{true}} = 0$ when $|R| > c_\alpha$. Under $H_0$ the placebo pool is uncontaminated and, under A1--A3, shares the first two moments of the deviation the test must span, so $c_\alpha$ is an approximately level-$\alpha$ critical value and reading the set as a confidence interval inverts a family of tests with this single value. The symmetric absolute-deviation construction is essential: the absolute value absorbs asymmetric location shifts into width, whereas constructions that center on the sample mean or use asymmetric quantiles move the confidence interval the wrong way under a genuine effect (coverage collapsed to 8--16\% at high intensity in exploratory runs).

\paragraph{Studentized variant.}
When placebos are scarce the symmetric confidence interval can be too coarse. A studentized variant---one that divides each bootstrap replicate by its own estimated standard error to normalize for scale---adds coverage by separating what the bootstrap is good at from what it is not: it borrows the \emph{shape} of the reference distribution from the placebo pool but rescales each replicate by its own internal standard error, then sets the half-width from the observed cross-unit standard error. The effect is to let the pool decide the quantile while the panel decides the scale. Concretely, let $r_i = \sum_{t \geq T_i^*} \hat{\Delta}_{i,t} \big/ \sum_{t \geq T_i^*} Y_{i,t}$ be the per-unit relative effect, $\widehat{\mathrm{SE}}_R = s_R/\sqrt{M}$ the observed cross-unit standard error with $s_R = \mathrm{sd}(r_1,\ldots,r_M)$, and $\widehat{\mathrm{se}}^{(m)}$ the internal standard error of the $m$-th bootstrap replicate (Appendix~\ref{app:method} gives the formula). Each replicate is standardized into $S^{(m)} = R^{(m)}/\widehat{\mathrm{se}}^{(m)}$, giving
\begin{equation}\label{eq:stud}
  \bigl[\,R - c_\alpha^S\,\widehat{\mathrm{SE}}_R,\; R + c_\alpha^S\,\widehat{\mathrm{SE}}_R\,\bigr], \quad
  c_\alpha^S = Q_{1-\alpha}\!\bigl(|S^{(1)}|, \ldots, |S^{(B)}|\bigr).
\end{equation}
In short, the bootstrap decides the \emph{shape} of the interval (the quantile $c_\alpha^S$) and the panel decides its \emph{size} (the half-width, set by $\widehat{\mathrm{SE}}_R$). We borrow only the shape and do not claim the studentized statistic follows any fixed distribution. This is exactly the right division of labor when the placebo pool gets the shape of the error distribution right but its overall magnitude wrong.\footnote{The studentized statistic does not converge to a fixed reference distribution as $M$ grows: when $\mathbb{E}[\bar{b}] \neq 0$ the ratio $\bar{b}/(s_R/\sqrt{M})$ diverges in $M$, which is why the variant borrows shape rather than claiming a fixed reference.}

\begin{remark}[Practical implementation]\label{rem:practical}
Two deployment choices are worth noting. First, the placebo pool is partitioned into $K$ strata on pre-treatment covariates so that bootstrapped cohorts match the treated composition; in production we further sub-stratify by a hash of the unit identifier, capping each unit's draws per replicate to ensure unit diversity. Second, when the median pre-treatment outcome is very small (below the 5th percentile), the denominator of the relative metric~\eqref{eq:aggregate} is unstable and we revert to the absolute-gap metric $G = \sum_i \hat{\Delta}_i$, to which the construction applies identically.
\end{remark}

\subsection{Assumptions and Coverage Guarantee}\label{sec:assumptions}
The bootstrap critical value $c_\alpha$ of the calibrated-at-zero construction (Eq.~\eqref{eq:ci}) is calibrated at the zero null---it measures how large the SC model's error can be in the absence of treatment---so the resulting confidence interval is valid whenever the placebo gaps faithfully represent the bias that contaminates the real estimate. Three conditions make this precise (more details in Appendix~\ref{app:assumptions}):
\begin{itemize}
\item[\textbf{A1.}] \textbf{Additive separability.} The unit-level gap decomposes as $\hat{\Delta}_i = \tau_i + b_i + \varepsilon_i$, with the model bias $b_i$ unaffected by the treatment itself.
\item[\textbf{A2.}] \textbf{Bias transportability across onsets.} The placebo gap at a backdated onset is drawn from the same bias distribution as the gap at the actual onset, so that placebo gaps are informative about the real error. This is the identification assumption: it can fail when deeper backdating degrades SC fit quality.
\item[\textbf{A3.}] \textbf{Cross-unit pool structure.} Placebo errors are approximately i.i.d.\ across units (up to weak donor-pool dependence), with across-shift location spread small relative to cross-unit variance (checkable in-sample).
\end{itemize}
Under A1--A3 the placebo pool is an unbiased sample of the SC model's error distribution conditional on the realized panel, and the bootstrap dispersion converges to the factor-orthogonal component of $\mathrm{Var}(R - R_{\mathrm{true}})$. The remaining factor-dependent component---variation that would require a fresh draw of the latent common factors---is structurally inaccessible from a single panel and produces the sub-nominal coverage ceiling observed in the simulations (see Appendix~\ref{app:ceiling}).

%% =========================================================
\section{Simulation Study}\label{sec:sim}

\paragraph{Data-generating process (DGP).}
We generate panel data from a latent factor model: $Y_{i,t} = \boldsymbol{\lambda}_i^\top \mathbf{f}_t + \gamma_i \mathbf{1}[t \geq T_i^*] + \varepsilon_{i,t}$, with $r$ common factors following AR(1) processes. Treated loadings are displaced by $\eta$ on the first factor, so simplex-constrained donor weights cannot match the treated exposure and induce a shared-sign bias. Treatment intensity is $\gamma = c\,\tilde{b}$ where $\tilde{b}$ is the median replication-level bias, reported as $\tau/\tilde{b}$; the baseline configuration uses $N_d = M = 50$, $T = 80$, $r = 3$, $\eta = 0.25$, 1000 replications ($R^2 \approx 0.955$; full specification in Appendix~\ref{app:dgp}).

\begin{figure}[!h]
\centering
\includegraphics[width=\linewidth]{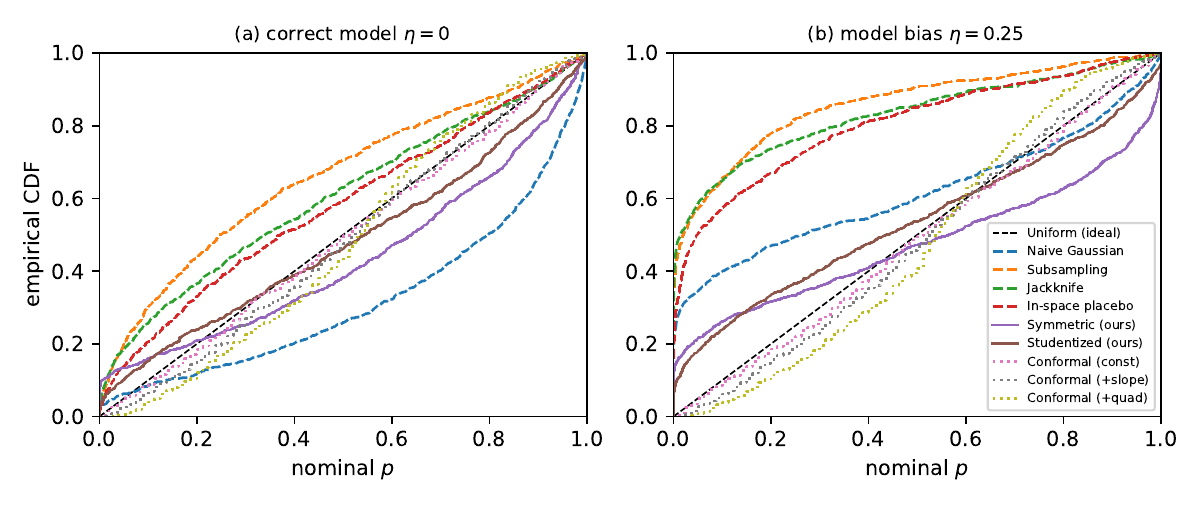}
\vspace{-0.3in}
\caption{Empirical CDF of null $p$-values against the Uniform$(0,1)$ diagonal (1000 replications). A calibrated test lies on the diagonal; a curve bowed above it over-rejects. Panel~(a): correctly specified model ($\eta=0$); panel~(b): systematic bias ($\eta=0.25$).}\label{fig:ppplot}
\end{figure}

\paragraph{Null calibration is necessary but not discriminating.}
Under loading displacement ($\eta = 0.25$) the symmetric bootstrap lifts null coverage from 0.655 to 0.786 (studentized: 0.821) at narrower width than the Gaussian (0.114 versus 0.152). The null $p$-value diagnostic (Figure~\ref{fig:ppplot}) confirms that variance- and permutation-based methods over-reject while both the proposed and conformal variants stay calibrated. Null calibration is shared with conformal; the discriminator is trajectory-agnosticism. When the SC model is correct ($\eta = 0$) the Gaussian confidence interval is near-nominal (0.937 versus the bootstrap's 0.866): the method is for the bias-dominated regime, not a default replacement.

\vspace{-0.1in}

\begin{figure*}[!h]
\centering
\includegraphics[width=\textwidth]{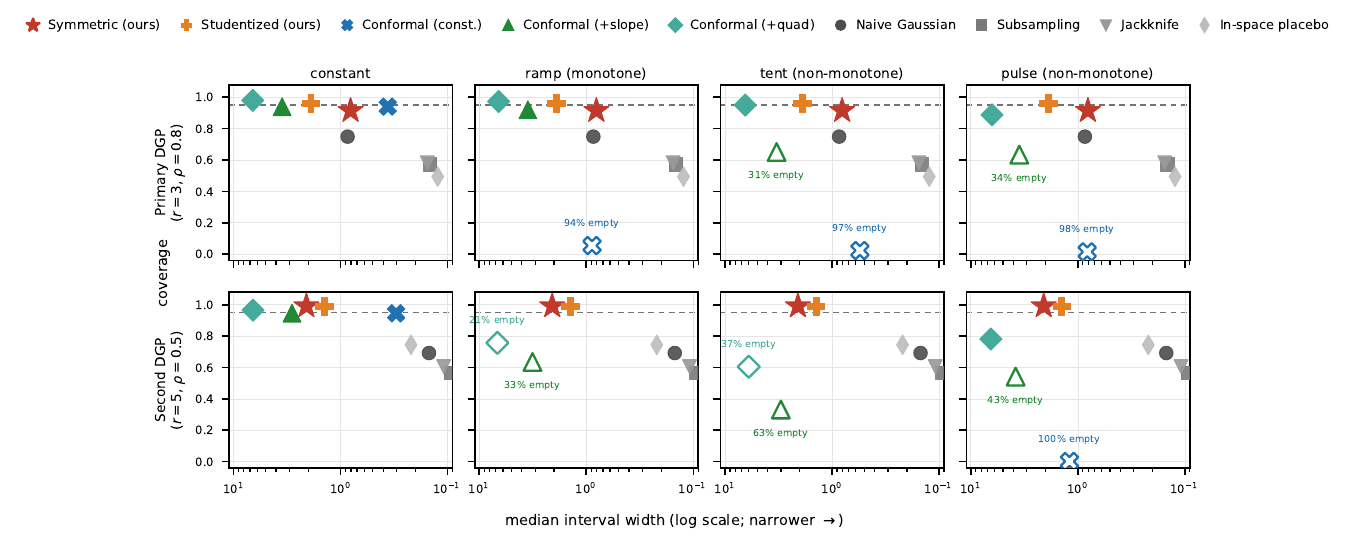}
\vspace{-0.35in}
\caption{Coverage versus median confidence interval width (log scale, flipped so narrower is right; top-right is desirable, dashed line is nominal 0.95) across four effect trajectories sharing the same mean $5\tilde{b}$. Each row is a DGP family: top $r=3$, $\rho=0.8$; bottom $r=5$, $\rho=0.5$. Hollow points indicate ${>}20\%$ empty intervals. Details in Appendix~\ref{app:fam2}.}\label{fig:traj}
\includegraphics[width=0.7\textwidth]{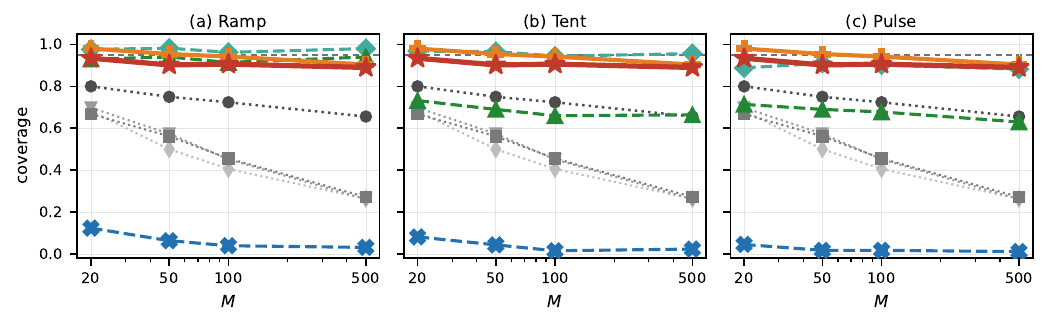}
\vspace{-0.25in}
\caption{Coverage versus treated-unit pool size $M$ under non-constant effect shapes (500 replications per cell, $\tau = 5\tilde{b}$, $\eta = 0.25$). The proposed variants (red, orange) hold nearly flat across $M$ in every panel. Non-conformal lines are identical across panels (trajectory-agnosticism, also shown in Figure~\ref{fig:traj}); conformal lines shift as the true shape departs from the assumed family. Dashed line is nominal 0.95.}\label{fig:msweep}
\end{figure*}
\vspace{-0.2in}

\paragraph{Trajectory-agnosticism is the decisive test.}\label{sec:regime}
We compare against moving-block conformal inference~\cite{ChernozhukovWuthrichZhu2021} applied to the aggregate treated series---its most favorable adaptation (details in Appendix~\ref{app:frontier}). On a constant shift conformal is stronger (0.965 versus 0.805), because constant displacement preserves residual exchangeability. The decisive test holds the mean effect fixed at $5\tilde{b}$ and varies only how it evolves over time (Figure~\ref{fig:traj}). The placebo bootstrap holds 0.91 coverage at fixed width across all four shapes. Conformal collapses off-constant: the constant family produces empty intervals in 94--98\% of samples, and richer families recover coverage only at ${\approx}8\times$ the width. No fixed conformal family is both flexible enough for an unknown shape and tight enough to be useful. A second DGP (Figure~\ref{fig:traj}, bottom row; $r = 5$, $\rho = 0.5$) sharpens the separation: even the quadratic family now under-covers non-monotone shapes (Appendix~\ref{app:fam2}). The prediction intervals of \citet{CattaneoFengTitiunik2021} are not competitive on this estimand (Appendix~\ref{app:frontier}).

\paragraph{Coverage stability across pool size.}\label{sec:msweep}
The baseline comparison fixes $M = 50$. To test whether the advantage persists as the treated-unit pool grows, we sweep $M \in \{20, 50, 100, 500\}$ with $\tau = 5\tilde{b}$ under three non-constant shapes (ramp, tent, pulse), 500 replications per cell, all nine methods (Figure~\ref{fig:msweep}). The symmetric placebo bootstrap holds coverage in the band $0.890$--$0.934$ across all twelve cells, nearly flat in $M$, while the Gaussian confidence interval falls from $0.80$ to $0.66$. Conformal separates by shape: the quadratic family covers well under ramp and tent but degrades under pulse, where at $M = 500$ its coverage ($0.882$) falls below the symmetric bootstrap's ($0.890$); no fixed conformal family stays above $0.89$ in every cell (full breakdown in Appendix~\ref{app:msweep}). Together with Figure~\ref{fig:traj}, the $M$-sweep confirms robustness across both dimensions a practitioner does not control: the temporal shape of the effect and the treated-unit pool size.

\vspace{-0.15in}
%% =========================================================
\section{Conclusion}\label{sec:discussion}

The placebo-in-time bootstrap is not a default replacement. Where the SC model is well specified the Gaussian confidence interval is better, and for a constant shift conformal inference is stronger. The method targets the intersection neither serves: a fixed estimator with suspected bias and an unknown temporal shape. Its differentiator is trajectory-agnosticism---valid coverage at fixed width with no assumptions about how the effect evolves---because it corrects a failure of bias, not variance.

Three caveats bound the guarantee. First, the coverage ceiling (Appendix~\ref{app:ceiling}): conditioning on a single panel misses the factor-dependent variance component, producing sub-nominal coverage (0.79--0.87). Under the null this ceiling deepens as $M$ grows, but under non-constant effects the $M$-sweep (Section~\ref{sec:msweep}) shows coverage stabilizes, so the method is not ``data-hungry'' in the regime it targets. Second, additive separability (A1): when treatment perturbs the treated loading, null coverage falls. Third, transportability (A2): when bias is a volatile nonlinear function of the factor path, backdated placebos no longer track it. Extending to staggered-onset designs with time-varying confounding is the main open problem.

\vspace{-0.15in}
%% =========================================================
\begin{acks}
We thank Jin Ye, Yael May-Rom, Ken Spencer, and Kevin Kallock for domain feedback, and our senior business leaders Michael Kiermaier and Stephen Moss for their partnership; Nan Wang for applying this method in new domains at Databricks; and Sam Shah, Feng Pan, and Divy Menghani for the Data Science leadership support.
\end{acks}

%% =========================================================
\bibliographystyle{ACM-Reference-Format}
\bibliography{references}

%% =========================================================
\appendix

\section{Extended Literature Survey}\label{app:litsurvey}
Placebo inference for synthetic control originates with Abadie and Gardeazabal~\cite{Abadie2003} and the in-space permutation test of Abadie et al.~\cite{Abadie2010}, which reassigns treatment to each donor and ranks the treated effect among the resulting placebos; Abadie~\cite{Abadie2021} surveys the methodology, and in the taxonomy of Eggers et al.~\cite{EggersTunonDafoe2024} our construction is a treatment placebo elevated from diagnostic to inference. The in-time placebo we build on was introduced by Abadie et al.~\cite{Abadie2015} as a binary diagnostic; Chen and Yan~\cite{ChenYan2023} combine in-time and in-space placebos but retain a rank-based $p$-value. In-space permutation inference has known fragilities: its symmetry assumption can fail~\cite{HahnShi2017}, heterogeneous pre-treatment fit distorts size~\cite{FermanPinto2021,FermanPinto2017}, resolution is limited by donor-pool size~\cite{LeiSudijono2025}, and Firpo and Possebom~\cite{FirpoPossebom2018} invert a sensitivity analysis to obtain confidence sets.

A second line bounds or estimates the error analytically rather than resampling it: the conformal and cross-fitted procedures of Chernozhukov et al.~\cite{ChernozhukovWuthrichZhu2021,ChernozhukovWuthrichZhu2025} and the prediction intervals of Cattaneo et al.~\cite{CattaneoFengTitiunik2021,CattaneoFengPalombaTitiunik2025}---the closest neighbor to our target, which we compare against directly in Section~\ref{sec:regime}. On the resampling side, the standard bootstrap is inconsistent for SC average effects~\cite{Li2020} and for matching estimators~\cite{AbadieImbens2008,OtsuRai2017}, a pathology we sidestep by resampling placebo effects---regular quantities unaffected by the constraint that SC weights must be non-negative and sum to one---rather than the SC weights. For many treated units under staggered adoption, Ben-Michael et al.~\cite{BenMichaelFellerRothstein2022}, Xu~\cite{Xu2017}, and Cao et al.~\cite{CaoLuWu2019} develop estimation and large-sample inference; synthetic difference-in-differences~\cite{Arkhangelsky2021} supplies unit-dimension variance estimators; and Bayesian approaches~\cite{Brodersen2015,PangLiuXu2022} quantify uncertainty under a posited generative model. Estimator-side correction such as augmented SC~\cite{BenMichaelFellerRothstein2021} reduces the bias at the estimation stage and is complementary to the inference-stage construction studied here.

%% =========================================================
\section{Assumptions and Coverage-Ceiling Derivation}\label{app:method}
This appendix gives the formal assumption statements referenced in Section~\ref{sec:assumptions}, defines the studentized inner scale, and derives the variance decomposition that explains the sub-nominal coverage ceiling.

\subsection{Detailed Formulation}\label{app:assumptions}

\begin{assumption}[Additive separability]\label{as:sep}
The unit-level gap decomposes as $\hat{\Delta}_i = \tau_i + b_i + \varepsilon_i$ (Eq.~\eqref{eq:decomposition}), with the model bias $b_i$ (the gap that would obtain under no treatment) unaffected by the treatment itself.
\end{assumption}

\begin{assumption}[Bias transportability across onsets]\label{as:transport}
Let $\mathcal{G}$ denote the $\sigma$-field (the information set) generated by the panel's common component (under the latent factor DGP of Section~\ref{sec:sim}, the realized sequence of common factors $\mathcal{G} = \sigma(\mathbf{f}_1, \ldots, \mathbf{f}_T)$); conditional on $\mathcal{G}$, the gaps remain random through the loading and noise draws. For every shift $\delta \leq \delta_{\max}$, the placebo gap at the backdated onset $T_i^* - \delta$ has the same cross-unit conditional variance given $\mathcal{G}$ as the gap at the actual onset $T_i^*$, and its conditional mean given $\mathcal{G}$ is the same function of the placebo window $[T_i^* - \delta,\, T]$ that the actual-onset mean is of $[T_i^*,\, T]$. The two windows overlap but differ, so the placebo locations track the actual-onset location only up to a window attenuation.
\end{assumption}

\begin{assumption}[Cross-unit pool structure]\label{as:weakdep}
Conditional on $\mathcal{G}$, the per-unit vectors of placebo errors across shifts are i.i.d.\ over units (up to a weak dependence of order $O(N_d^{-1})$ induced by sharing the same donor pool) with finite fourth moments; write $\sigma^2(\mathcal{G})$ for the common cross-unit conditional variance and $s_\mu^2(\mathcal{G})$ for the across-shift spread of the $\mathcal{G}$-conditional placebo locations. The spread is small relative to the cross-unit variance, $s_\mu^2(\mathcal{G}) \leq \kappa\, \sigma^2(\mathcal{G})$ for some constant $\kappa < 1$; since $s_\mu^2(\mathcal{G})$ is estimable from the across-shift means of the realized pool, this bound is checkable in-sample. No bound is placed on within-unit cross-shift correlations.
\end{assumption}

\paragraph{Studentized inner scale.}
The internal standard error of the $m$-th replicate in Eq.~\eqref{eq:stud} is the mean-of-ratios standard error over the per-draw ratios,
\begin{equation}\label{eq:innerse}
  \widehat{\mathrm{se}}^{(m)} = \frac{\mathrm{sd}\bigl(\{r^{(m)}_j\}_{j \in \mathcal{S}^{(m)}}\bigr)}{\sqrt{|\mathcal{S}^{(m)}|}}, \qquad
  r^{(m)}_j = \frac{\sum_{t \geq T_i^{(\delta)}} \hat{\Delta}^{(\delta)}_{i,t}}{\sum_{t \geq T_i^{(\delta)}} Y_{i,t}},
\end{equation}
where $j = (i,\delta)$ indexes the pairs drawn into the replicate multiset $\mathcal{S}^{(m)}$.

\subsection{Coverage-ceiling variance decomposition}\label{app:ceiling}
At the null ($\tau_i = 0$, so $R_{\mathrm{true}} = 0$) the SC gap under the factor DGP of Section~\ref{sec:sim} is $\hat{\Delta}_{i,t} = \mathbf{d}_i^\top \mathbf{f}_t + e_{i,t}$, with loading mismatch $\mathbf{d}_i = \boldsymbol{\lambda}_i - \sum_j w_{ij} \boldsymbol{\lambda}_j$ and noise gap $e_{i,t} = \varepsilon_{i,t} - \sum_j w_{ij}\varepsilon_{j,t}$. Writing $\bar{\mathbf{F}} = \sum_{t \geq T^*} \mathbf{f}_t$, $\bar{e}_i = \sum_{t \geq T^*} e_{i,t}$, and $D_0$ for the expected per-unit post-window outcome sum, linearizing $R$ around $M D_0$ gives $R \approx \bar{u} = \frac{1}{M}\sum_i u_i$ with $u_i = (\mathbf{d}_i^\top \bar{\mathbf{F}} + \bar{e}_i)/D_0$.

\begin{proposition}[Heuristic variance decomposition]\label{prop:decomp}
Under Assumptions~\ref{as:sep}--\ref{as:weakdep} and the linearization above, with $M$ and $\delta_{\max}$ fixed:

\emph{(i)} Across panel redraws the error variance decomposes by the law of total variance as
\begin{equation}\label{eq:totvar}
  \mathrm{Var}(R - R_{\mathrm{true}}) \;=\; \underbrace{\mathbb{E}\bigl[\mathrm{Var}(R \mid \mathcal{G})\bigr]}_{\text{factor-orthogonal}} \;+\; \underbrace{\mathrm{Var}\bigl(\mathbb{E}[R \mid \mathcal{G}]\bigr)}_{\text{factor-dependent}},
\end{equation}
where the factor-orthogonal component is the variation due to loading and noise draws for a fixed realization of the common factors, and the factor-dependent component is the variation due to different realizations of those factors. Under the additional approximation $\mathbb{E}[\mathbf{d}_i \mid \mathcal{G}] = \mathbb{E}[\mathbf{d}_i]$ (the expected loading mismatch does not depend on the realized factor path), the factor-dependent component equals $\mathbb{E}[\mathbf{d}_i]^\top \mathrm{Var}(\bar{\mathbf{F}})\,\mathbb{E}[\mathbf{d}_i]/D_0^2$, strictly positive whenever displacement makes $\mathbb{E}[\mathbf{d}_i] \neq \mathbf{0}$ and $\mathrm{Var}(\bar{\mathbf{F}})$ is positive definite (as under AR(1) factors).

\emph{(ii)} Fixing the realized panel, as $B \to \infty$ the bootstrap dispersion $\mathrm{Var}(R^{(m)} \mid \mathrm{panel})$ converges to the pool variance divided by $M$, whose $\mathcal{G}$-conditional expectation is $[\sigma^2(\mathcal{G}) + s_\mu^2(\mathcal{G})]/M + O(M^{-2})$: the bootstrap estimates the realized factor-orthogonal scale and retains only the attenuated trace $s_\mu^2(\mathcal{G})$ of the factor-dependent component.
\end{proposition}

\emph{Heuristic derivation.} Conditional on $\mathcal{G}$ the pairs $(\mathbf{d}_i, \bar{e}_i)$ are i.i.d.\ across units (up to $O(N_d^{-1})$ donor-pool dependence), so the law of total variance gives (i) exactly for $\bar{u}$; the factor-dependent term uses the mean-invariance approximation $\mathbb{E}[\mathbf{d}_i \mid \mathcal{G}] = \mathbb{E}[\mathbf{d}_i]$, since the SC weights are fit on pre-window outcomes that contain the realized factors. For (ii), a uniformly drawn pool element has $\mathcal{G}$-conditional marginal variance $\sigma^2(\mathcal{G}) + s_\mu^2(\mathcal{G})$ (the locations share the final $L$ periods, so their spread $s_\mu^2(\mathcal{G})$ is an attenuated trace), and i.i.d.\ pair resampling of size $M$ converges to $\widehat{\mathrm{Var}}_{\mathrm{pool}}(u)/M$ with that conditional expectation. The delta-method linearization, mean-invariance approximation, and bootstrap consistency for the smooth ratio functional are all heuristic; we do not claim an exact coverage formula, so this is a conjecture rather than a coverage theorem. \hfill$\square$

Empirically the omitted factor-dependent component is a null-coverage shortfall of about 8.5pp under the $\eta = 0$ DGPs and 16.4pp under loading displacement (1000 replications), and a parametric panel bootstrap that redraws the common factors and noise restores precisely this component (null coverage ${\approx}0.80 \to 0.92$--$0.99$), consistent with the decomposition but not confirming it---the parametric bootstrap targets the unconditional error variance, a different and larger quantity. The ceiling is therefore the price of conditioning on a single panel, not a defect of the estimator; it deepens as treated units grow under the null ($M$-sweep at $\tau = 0$, $500$ replications per cell: symmetric null coverage $0.820 \to 0.780 \to 0.720 \to 0.650$ from $M = 20$ to $500$, the Gaussian confidence interval falling in parallel $0.746 \to 0.656 \to 0.608 \to 0.552$), a milder version of the shrink-around-bias pathology the method corrects in the Gaussian confidence interval. Crucially, this null-regime degradation does not extend to the setting the method targets: under non-constant effects at $\tau = 5\tilde{b}$ (Section~\ref{sec:msweep}), the symmetric variant's coverage is nearly flat across the same $M$ range ($0.890$--$0.934$, range $4.4$ percentage points), confirming that the ``data-hungry'' appearance under the null is an artifact of testing in the wrong regime.

\section{Simulation Details and Robustness}\label{app:sim}
This appendix specifies the full DGP, defines every benchmark method reported in the experiments, and checks trajectory-agnosticism under a second DGP.

\subsection{Primary DGP and configuration}\label{app:dgp}
The outcome model is $Y_{i,t} = \boldsymbol{\lambda}_i^\top \mathbf{f}_t + \gamma_i \mathbf{1}[t \geq T_i^*] + \varepsilon_{i,t}$, where $\boldsymbol{\lambda}_i \in \mathbb{R}^r$ is the loading vector, $\mathbf{f}_t$ are common factors following independent AR(1) processes with persistence $\rho$, $\gamma_i$ is the per-period effect, and $\varepsilon_{i,t}$ is mean-zero noise. Donor loadings are $\mathrm{Uniform}(0,1)$; treated loadings are drawn identically then displaced on the first factor by $\eta$, so simplex-weighted donor combinations undermatch the treated factor exposure and induce a bias $b_i$ shared in sign across treated units. Treatment intensity is calibrated to the model's intrinsic error: null replications define $\tilde{b}$ as the median absolute replication-level bias of $R$, and the per-period effect is $\gamma = c\,\tilde{b}$ for $c \in \{0, 0.5, 1, 2, 5, 10\}$, reported as $\tau/\tilde{b}$. The baseline configuration is $N_d = M = 50$, $T = 80$, $T^* = 61$ ($L = 20$), $r = 3$ factors, $\rho = 0.8$, $\sigma_\varepsilon = 0.30$, $\eta = 0.25$, $\delta_{\max} = 30$, $B = 500$, $K = 3$ strata. Each replication generates a fresh panel, fits SC by constrained least squares on the simplex, and records whether each confidence interval covers $R_{\mathrm{true}}$; headline tables use 1000 replications (binomial SE $\le 1.5$pp).

\subsection{Benchmark methods}\label{app:frontier}
Every method reported in the experiments is defined here. All operate on the same per-unit gaps $\hat\Delta_i = \tau_i + b_i + \varepsilon_i$ of Eq.~\eqref{eq:decomposition} and target the aggregate $R$ of Eq.~\eqref{eq:aggregate} (or, equivalently in the equal-size simulations, $\bar\Delta = M^{-1}\sum_i \hat\Delta_i$).

\emph{Naive Gaussian confidence interval.} The interval $\bar\Delta \pm z_{\alpha/2}\,\hat\sigma_\Delta/\sqrt{M}$ of Eq.~\eqref{eq:naive-ci}, with $\hat\sigma_\Delta^2 = (M-1)^{-1}\sum_i (\hat\Delta_i - \bar\Delta)^2$. It is calibrated for $\varepsilon_i$ and blind to $\bar b$.

\emph{Jackknife.} The delete-one-unit ratio variance. With $\bar\Delta_{(-i)}$ the leave-one-out mean and $\bar\Delta_{(\cdot)} = M^{-1}\sum_i \bar\Delta_{(-i)}$,
\begin{equation}\label{eq:jack}
  \hat{V}_{\mathrm{jack}} = \frac{M-1}{M}\sum_{i=1}^{M}\bigl(\bar\Delta_{(-i)} - \bar\Delta_{(\cdot)}\bigr)^2,
\end{equation}
giving $\bar\Delta \pm z_{\alpha/2}\sqrt{\hat V_{\mathrm{jack}}}$. Because deletion does not move the shared bias $\bar b$, $\hat V_{\mathrm{jack}}$ estimates dispersion about the contaminated center and coincides with the Gaussian confidence interval in the figures.

\emph{Subsampling}~\cite{Li2020}. Draw subsamples of size $\ell = \lceil M^{0.66}\rceil$ without replacement; for each, form the recentered statistic $\sqrt{\ell}\,(\bar\Delta^{(\ell)} - \bar\Delta)$, and take the empirical $\alpha/2$ and $1-\alpha/2$ quantiles of its distribution $L_\ell$ to invert
\begin{equation}\label{eq:subsample}
  \Bigl[\,\bar\Delta - \tfrac{1}{\sqrt{M}} L_\ell^{-1}(1-\tfrac\alpha2),\;\; \bar\Delta - \tfrac{1}{\sqrt{M}} L_\ell^{-1}(\tfrac\alpha2)\,\Bigr].
\end{equation}
The Politis--Romano rate $\ell/M \to 0$ is what restores consistency where the full-sample bootstrap fails~\cite{Li2020}; recentering on $\bar\Delta$ leaves $\bar b$ in the center.

\emph{In-space placebo}~\cite{Abadie2010}. Leave-one-donor-out: reassign treatment to each donor $g$ in the pool $\mathcal{D}$, compute its placebo aggregate $R^{[g]}$ by refitting SC with $g$ held out as pseudo-treated, and form the donor-permutation band from the empirical quantiles of $\{R^{[g]}\}_{g\in\mathcal{D}}$,
\begin{equation}\label{eq:inspace}
  \Bigl[\,Q_{\alpha/2}\bigl(\{R^{[g]}\}\bigr),\;\; Q_{1-\alpha/2}\bigl(\{R^{[g]}\}\bigr)\,\Bigr],
\end{equation}
with the two-sided $p$-value the rank of $|R|$ among $\{|R^{[g]}|\}$. Validity requires treated--donor exchangeability, which structural mismatch violates.

\emph{Conformal (moving-block).} Following \citet{ChernozhukovWuthrichZhu2021}, to test $H_0:\theta = \theta_0$ over the post-period, subtract the hypothesized effect, refit on the adjusted series, collect residuals $\hat u_t$, and form the block statistic $S(\theta_0) = \bigl(\sum_{t \ge T^*} \hat u_t^2\bigr)^{1/2}$. Cyclically permuting residuals in moving blocks of length $L$ gives the permutation reference $\{S_\pi\}$ and the $p$-value $p(\theta_0) = |\{\pi : S_\pi \ge S(\theta_0)\}|/|\Pi|$; inverting over a parametric family describing how the effect evolves over time, $\theta_k = a + bk + ck^2$ ($k = t-T^*$, degree $d\in\{0,1,2\}$), yields the confidence interval. The permutation count $|\Pi| = T$ floors the resolution at $1/T$.

\emph{Prediction intervals (\texttt{scpi})}~\cite{CattaneoFengTitiunik2021}. For each post-period $k$ the confidence interval is the SC point gap plus a two-part bound, $\hat\Delta_{T^*+k} \pm \bigl(E_{1,k} + E_{2,k}\bigr)$, where $E_{1,k}$ bounds the in-sample weight-estimation error (from the constrained-least-squares optimization, via the package's conditional simulation of $\hat w$) and $E_{2,k}$ is a quantile of the out-of-sample error $e_{T^*+k}$ under a posited noise class. We run it through the authors' \texttt{scpi} package on the aggregate series.

\paragraph{Implementation notes.}
We run conformal on the aggregate treated series rather than per unit---the most favorable adaptation, since pooling $M$ units smooths the full-sample residuals toward exchangeability. For each degree $d$, the nuisance coefficients of the parametric family are profiled out by a max-$p$ search over an $11$-point slope grid and (for the quadratic) a $5$-point curvature grid, then the endpoints are found by bisection; the fully flexible per-period family leaves the interval vacuous. Reported coverage is usable coverage, with empty intervals counted as non-coverage. For \texttt{scpi}, native per-period coverage is 0.50--0.62 and joint (simultaneous) coverage 0.02--0.13 on this estimand. A summed-band aggregate appears to reach 1.00, but only under \emph{our} aggregation of the per-period bands rather than the authors' intended construction, so we report the native numbers and exclude the summed-band figure from any headline comparison.

\subsection{$M$-sweep: full results}\label{app:msweep}
Under the three non-constant shapes (ramp, tent, pulse) at $\tau = 5\tilde{b}$, the six non-conformal methods produce identical coverage across shapes (trajectory-agnosticism), so only the $M$ dimension matters: the symmetric bootstrap stays in $0.890$--$0.934$ (range $4.4$ percentage points), the studentized in $0.904$--$0.980$ ($7.6$ percentage points), the Gaussian drops from $0.80$ to $0.66$ ($14.4$ percentage points), and subsampling ($0.27$--$0.67$), jackknife ($0.26$--$0.70$), and in-space placebo ($0.26$--$0.68$) degrade faster still. The conformal families separate by shape: constant collapses to $0.012$--$0.124$ everywhere; slope recovers under ramp ($0.91$--$0.94$) but fails under tent ($0.66$--$0.73$) and pulse ($0.63$--$0.71$); quadratic covers well under ramp ($0.96$--$0.98$) and tent ($0.94$--$0.97$) but under pulse drops to $0.882$--$0.910$, falling below the symmetric bootstrap at $M = 500$.

\subsection{Robustness under a second DGP family}\label{app:fam2}
To check that trajectory-agnosticism is not an artifact of the primary three-factor structure, we re-run the decisive comparison of Section~\ref{sec:regime} under a materially different DGP: $r = 5$ factors, factor persistence $\rho = 0.5$ (versus 0.8 in the primary configuration), and loading displacement $\eta = 0.35$, with the mean per-period effect held fixed at $5\tilde{b}$ across the four trajectories (constant, ramp, tent, pulse). The displacement is a stationary loading offset, so bias transportability (A2) holds. The bottom row of Figure~\ref{fig:traj} reports coverage versus median width by trajectory. The symmetric placebo bootstrap stays trajectory-invariant (0.99 coverage across all four shapes), confirming the primary finding under a different factor geometry. Conformal again fails off-constant: the constant family is empty in 100\% of samples on the non-constant trajectories, and---the additional finding relative to the primary DGP---even the quadratic family, which recovered all four shapes there, now under-covers the non-monotone effects (0.60--0.78). The richer factor structure thus \emph{sharpens} the separation: the functional form a practitioner would need is both unknown and, at fixed flexibility, no longer sufficient.
\end{document}